\documentclass[english,pra,preprint,aps,floats]{revtex4}
\usepackage[T1]{fontenc}
\usepackage[latin1]{inputenc}

\usepackage{graphicx}
\usepackage{amsmath}
\usepackage{babel}
\usepackage{amssymb}
\providecommand{\U}[1]{\protect\rule{.1in}{.1in}}

\begin{document}

\title{Threshold Bound States}

\author{W. A. Berger
\email{E-mail: wb@wberger.com}, H. G. Miller
\email{E-mail: hmiller@maple.up.ac.za}}
\affiliation{Department of Physics, University of Pretoria, Pretoria 0002, South
Africa}
\author{and D. Waxman\email{D.Waxman@sussex.ac.uk}}
\affiliation{School of Life Sciences, The University of Sussex, Brighton, Sussex
BN1 9QH, ,
Sussex, UK}

\begin{abstract}
Relationships between the coupling constant and the binding energy of
threshold bound states are obtained in a simple manner from an iterative
algorithm for solving the eigenvalue problem. The absence of threshold bound
states in higher dimensions can be easily understood.

\noindent PACS\ 03.65.Ge,\ 02.60.Lj

\end{abstract}
\maketitle

In a mathematically elegant paper, Simon studied the one and two dimensional
Schrodinger operators $-\partial^{2}/\partial x^{2}+\lambda V(x)$ and
$-\bigtriangleup+\lambda V(\mathbf{x})$ where either $V(x)$ or $V(\mathbf{x})
$ is described in this work as the potential and the parameter $\lambda$
($>0$) is termed the strength of the potential. Simon provided necessary and
sufficient conditions for the existence of a bound state when $\lambda$ is
small\cite{S76}. In one dimension a threshold bound state (i.e., one just
bound) exists for many finite short-range potentials and its binding energy is
an analytical function of $\lambda$\cite{F70,Z87}. Furthermore, using the
theory of trace class determinants \cite{DS63,GK69}, a simple expansion for
the binding energy of the threshold bound state has been obtained(see Ref\cite{S76}).
More recently, Gat and Rosenstein have pointed out that
perturbative methods provide a suitable means for calculating the binding
energy of this state\cite{Gr93}. This is somewhat peculiar since a convergent
expansion for the binding energy in $\lambda$ exists, but no apparent poles
appear in the expansion of the S matrix to any finite order in perturbation
theory. Rather than use perturbation theory, we wish to point out that the
expression for the binding energy of the threshold bound state obtained by
Simon in one and two dimensions 
can easily be obtained from a simple non-perturbative iterative
algorithm\cite{W98} and we provide an intuitive explanation of the form of the
expansion in different dimensions.

In the algorithmic approach, eigenvalues and the associated eigenfunctions are
determined as functions of the strength of the potential, $\lambda$. To
illustrate the method, we consider the one-dimensional eigenvalue
equation\cite{W98}
\begin{equation}
\lbrack-\partial_{x}^{2}-\lambda V(x)]u(x)=-\epsilon u(x) \label{heq}%
\end{equation}
subject to%
\begin{equation}
\lim_{|x|->\infty}u(x)=0.
\end{equation}
Here $\partial_{x}=\partial/\partial x$, $\lambda>0$ and $\int V(x)dx\geq0 $. We
shall always assume $V(x)\rightarrow0\,$as$\,|x|\rightarrow\infty$. The energy
eigenvalue, $-\epsilon$ (with $\epsilon>0$), is negative and corresponds to a
bound state. Using Green's method a solution to equation (\ref{heq}) is given
by
\begin{equation}
u(x)=\lambda\int_{-\infty}^{\infty}G_{\epsilon}(x-x^{\prime})V(x^{\prime
})u(x^{\prime})dx^{\prime} \label{G}%
\end{equation}
where the Green's function $G_{\epsilon}(x)$ satisfies
\begin{equation}
\lbrack-\partial_{x}^{2}+\epsilon]G_{\epsilon}(x)=\delta(x) \label{gf}%
\end{equation}%
\begin{equation}
\lim_{|x|->\infty}G_{\epsilon}(x)=0.
\end{equation}
Normalizing $u(x)$ at an arbitrary $x_{ref}$
\begin{equation}
u(x_{ref})=1
\end{equation}
allows $\lambda$ to be written as (see Eq. (\ref{G}))
\begin{equation}
\lambda=\frac{1}{\int_{-\infty}^{\infty}G_{\epsilon}(x_{ref}-x^{\prime
})V(x^{\prime})u(x^{\prime})dx^{\prime}} \label{l}%
\end{equation}
which can then be used to eliminate $\lambda$ from equation (\ref{G}):%
\begin{equation}
u(x)=\frac{\int_{-\infty}^{\infty}G_{\epsilon}(x-x^{\prime})V(x^{\prime
})u(x^{\prime})dx^{\prime}}{\int_{-\infty}^{\infty}G_{\epsilon}(x_{ref}%
-x^{\prime})V(x^{\prime})u(x^{\prime})dx^{\prime}}. \label{u}%
\end{equation}
Using equations (\ref{l}) and (\ref{u}), $\lambda$ can then be determined as a
function of $\epsilon$ as follows. For a particular choice of $\epsilon$ Eq.
(\ref{u}) can be iterated from a reasonable starting point, $u_{0}(x)$:%
\begin{equation}
u_{n+1}(x)=\frac{\int_{-\infty}^{\infty}G_{\epsilon}(x-x^{\prime})V(x^{\prime
})u_{n}(x^{\prime})dx^{\prime}}{\int_{-\infty}^{\infty}G_{\epsilon}%
(x_{ref}-x^{\prime})V(x^{\prime})u_{n}(x^{\prime})dx^{\prime}} \label{iu}%
\end{equation}
until it converges and $\lambda$ can then be determined from Eq. (\ref{l}).
Repeating this procedure for a different value of $\epsilon$ yields a
different value of the strength of the potential, $\lambda$. When enough
$(\epsilon,\lambda)$ pairs have been determined, a simple interpolation
procedure can be used to determine the dependence of $\epsilon$ on $\lambda$.
Furthermore, for larger values of $\epsilon$ a simple relationship between
$\lambda$ and $\epsilon$ can be obtained for non-singular symmetric potentials
which vanish asymptotically which can be used to make the the algorithm more
efficient\cite{BM06}.

For small values of $\epsilon$, corresponding to states on the threshold of
being bound, the analytical dependence of $\epsilon$ on $\lambda$ may be
obtained approximating the Green's function, which satisfies Eq. (\ref{gf})
and is given by
\begin{equation}
G_{\epsilon}(x)=\frac{e^{-\sqrt{\epsilon}|x|}}{2\sqrt{\epsilon}}. \label{gf1}%
\end{equation}
Expanding $G_{\epsilon}(x)$ in $\epsilon$:%
\begin{equation}
G_{\epsilon}(x)=\frac{1}{{2\sqrt{\epsilon}}}+\ldots
\end{equation}
and substituting this into Eq. (\ref{u}) yields
\begin{equation}
u(x)=1+\ldots.
\end{equation}
to leading order in $\epsilon$. From Eq. (\ref{l}) one therefore easily
obtains the following approximate relationship between the coupling constant
$\lambda$ and $\epsilon$
\begin{equation}
\lambda=\frac{2\sqrt{\epsilon}}{\int_{-\infty}^{\infty}V(x)dx} \label{lamv}%
\end{equation}
which is valid for small values of $\epsilon$. For arbitrary small coupling it
provides an analytical expression for the lowest bound state for a large class
of potentials in one dimension provided $\int_{-\infty}^{\infty}V(x)dx\geq0$\cite{S76}.
Furthermore, had we included the standard factor of $1/2$ in the first term of
the eigenvalue equation, Eq. (\ref{heq}), then we would obtain%
\begin{equation}
\epsilon=\frac{1}{2}\lambda^{2}\left(  \int_{-\infty}^{\infty}V(x)dx\right)
^{2}%
\end{equation}
which is precisely the result obtained by Simon to $O(\lambda^{2})$.

An important feature of Eq. (\ref{G}) is that it has a solution for $\epsilon
$, for arbitrarily small $\lambda$. This follows since the Green's function,
Eq. (\ref{gf1}), is unbounded from above as $\epsilon\rightarrow0 $:%
\begin{equation}
\lim_{\epsilon\rightarrow0_{+}}G_{\epsilon}(x)=+\infty.
\end{equation}
Hence for very small values of $\lambda$, the value of $\epsilon$ can always
be adjusted until the product $\lambda G_{\epsilon}(x)$ is non-negligible.

We note these results can be generalized to higher dimensions since it can
easily be seen that Eqs (\ref{G}) and (\ref{u}) become
\begin{equation}
u(\mathbf{x})=\lambda\int_{all\ space}G_{\epsilon}(\mathbf{x}-\mathbf{x}%
^{\prime})V(\mathbf{x}^{\prime})u(\mathbf{x}^{\prime})d^{n}x^{\prime}
\label{Gn}%
\end{equation}
and
\begin{equation}
u(\mathbf{x})=\frac{\int_{all\ space}G_{\epsilon}(\mathbf{x}-\mathbf{x}%
^{\prime})V(\mathbf{x}^{\prime})u(\mathbf{x}^{\prime})d^{n}x^{\prime}}%
{\int_{all\ space}G_{\epsilon}(\mathbf{x}_{ref}-\mathbf{x}^{\prime
})V(\mathbf{x}^{\prime})u(\mathbf{x}^{\prime})d^{n}x^{\prime}}. \label{un1}%
\end{equation}
The essential difference arises only from the different form the Green's
function takes in different dimensions.

In two dimensions%
\begin{equation}
G_{\epsilon}(\mathbf{x})=\frac{1}{2\pi}K_{0}(\epsilon|\mathbf{x}|)
\end{equation}
where $K_{0}(\cdot)$ is a Bessel function of the second kind of order
zero\cite{AS65}. Expanding $K_{0}(\epsilon|\mathbf{x}|)$ for small $\epsilon$
yields%
\begin{equation}
K_{0}(\epsilon|\mathbf{x}|)=\ln\left(  1/\epsilon\right)  +\ln\left(
2e^{-\gamma}/|\mathbf{x}|\right)  +O(\epsilon^{2})
\end{equation}
where $\gamma=0.57721...$ is Euler's constant. Thus from Eq. (\ref{un1}) we
obtain, for sufficiently small $\epsilon$,%
\begin{equation}
u(\mathbf{x})=1+\ldots
\end{equation}
and therefore
\begin{equation}
\lambda\simeq\frac{1}{\ln\left(  \dfrac{1}{\epsilon}\right)  \dfrac{1}{2\pi}%
{\displaystyle\int}
V(\mathbf{x})d^{2}x} \label{lamv2}%
\end{equation}
As in one dimension 
provided $
{\displaystyle\int}
V(\mathbf{x})d^{2}x\geq0$,
 a threshold bound state exists for arbitrarily small $\lambda$ 
if ${\displaystyle\int} |V(\mathbf{x})|^{1+\beta} d^2x < \infty$ (some $\beta>0$) and
${\displaystyle\int} (1+x^2)^\beta |V(\mathbf{x})| d^2x < \infty$\cite{S76}. 
Again it is the divergence of the Green's function, at
fixed spatial argument, when $\epsilon\rightarrow0$, that leads to a threshold
bound state at arbitrarily small $\lambda$.

On the other hand in three dimensions the Green's function is
\begin{equation}
G_{\epsilon}(\mathbf{x})=\frac{e^{-\epsilon|\mathbf{x}|}}{4\pi|\mathbf{x}|}.
\label{gf3}%
\end{equation}
At fixed $|\mathbf{x}|$, this does not diverge as $\epsilon\rightarrow0$. This
is suggestive of the known fact that in three and higher dimensions an
arbitrarily weak attractive potential does not possess a bound state\cite{S76}; there
has to be a certain strength of the potential before it can support a bound
state. We note that the leading term in an expansion of $\epsilon$ of the
Green's function, in $n=3$ and higher dimensions, is not independent of
$\mathbf{x}$. This is different to the corresponding behaviour of the Green's
functions when $n=1$ and $n=2$ and suggests that $n=1$ and $n=2$ which may be
thought of as being atypical of all other dimensions.

To understand the property of an arbitrarily weak attractive potential to bind
a particle in $n=1$ and $n=2$ two dimensions but not in three or more
dimensions, we can relate it to an apparently different problem of how much
time a random walk in $n$ dimensions spends in the vicinity of its starting
position. Using Dirac notation in the general $n$-dimensional case where
$\mathbf{\hat{p}}$ is the momentum operator and $|\mathbf{x}\rangle$
($\langle\mathbf{x}|$) is an eigenket (eigenbra) of the coordinate operator,
we have%
\begin{equation}
G_{\epsilon}(\mathbf{x})=\langle\mathbf{x}|(\mathbf{\hat{p}}^{2}%
+\epsilon)^{-1}|\mathbf{0}\rangle=\int_{0}^{\infty}dt\langle\mathbf{x}%
|e^{-\left(  \mathbf{\hat{p}}^{2}+\epsilon\right)  t}|\mathbf{0}\rangle.
\end{equation}
This quickly leads to
\begin{equation}
\lim_{\epsilon\rightarrow0}G_{\epsilon}(\mathbf{x})=%
{\displaystyle\int_{0}^{\infty}}
\frac{e^{-|\mathbf{x|}^{2}\mathbf{/(}4t\mathbf{)}}}{(4\pi t)^{n/2}}dt.
\end{equation}
In a formulation of the random walk in discrete space and discrete
time\cite{ID89}, it is precisely the analogue of $\lim_{\epsilon\rightarrow
0}G_{\epsilon}(\mathbf{x})$ which determines the mean time a random walk
spends in the vicinity of a site at position $\mathbf{x}$, given it was at
position $\mathbf{0}$ at time $t=0$. In $n=1$ and $n=2$ dimensions
$\lim_{\epsilon\rightarrow0}G_{\epsilon}(\mathbf{x})$ is infinite, implying an
infinite amount of time is spent at $\mathbf{x}$. By contrast, when $n\geq3$,
$G_{0}(\mathbf{x})$ is finite. In the continuous space quantum mechanical
problem considered here, it is precisely the finiteness (or lack of
finiteness) of $\lim_{\epsilon\rightarrow0}G_{\epsilon}(\mathbf{x})$ that
determines the dimensionalities where an arbitrarily weak potential can
possess a threshold bound state.

We conclude by pointing out that at least for small values of $\lambda$,
good approximate analytical relationships between $\epsilon$ and $\lambda$  
  exist in one and two dimensions which may be
used to improve the convergence rate of the aforementioned iterative
algorithm\cite{W98}.


\end{document}